\documentclass[aps,prl,preprint,superscriptaddress]{revtex4-2}
\usepackage[dvipdfmx]{graphicx}
\usepackage{amsmath,amssymb,bm}
\usepackage{caption}
\usepackage[version=3]{mhchem}
\usepackage{braket}
\usepackage{tabularx}
\newcolumntype{Y}{&gt;{\centering\arraybackslash}X}
\newcommand{\ground}{^{3}\mathrm{P}} 
\newcommand{\excited}{^{3}\mathrm{D}} 
\newcommand{\intermediate}{^{3}\mathrm{P}^{\mathrm{o}}} 
\newcommand{\um}{\mathrm{\mu m}} 
\newcommand{\us}{\mathrm{\mu s}} 
\newcommand{\pcc}{\mathrm{cm^{-3}}} 

\begin{document}
\title{Buffer gas cooling of carbon atoms}
\author{Takashi~Sakamoto}
\affiliation{Department of Applied Physics, Graduate School of Engineering, The University of Tokyo, 7-3-1 Hongo, Bunkyo-ku, Tokyo 113-8656, Japan}
\author{Kohei Suzuki}
\affiliation{Department of Applied Physics, Graduate School of Engineering, The University of Tokyo, 7-3-1 Hongo, Bunkyo-ku, Tokyo 113-8656, Japan}
\author{Kosuke~Yoshioka}
\email[Electronic address: ]{yoshioka@fs.t.u-tokyo.ac.jp}
\affiliation{Department of Applied Physics, Graduate School of Engineering, The University of Tokyo, 7-3-1 Hongo, Bunkyo-ku, Tokyo 113-8656, Japan}
\affiliation{Photon Science Center, Graduate School of Engineering,
  The University of Tokyo, 2-11-16 Yayoi, Bunkyo-ku, Tokyo 113-8656, Japan}

\date{\today}
\begin{abstract}
We demonstrate buffer gas cooling of carbon atoms to cryogenic temperatures. By employing pulsed two-photon excitation followed by vacuum ultraviolet fluorescence detection, we measured the arrival time distribution of the ablated carbon atoms to the detection volume at various helium buffer gas densities. The experimental data, corroborated by Monte Carlo simulations, reveal a rapid decrease in the local temperature of the carbon atom gas to approximately 10~K within tens of microseconds. The findings establish a major step towards novel research utilizing cold and ultracold carbon atoms.
\end{abstract}

\maketitle
Cold atoms are versatile tools for exploring science in diverse fields, such as observation of quantum degeneracy~\cite{anderson1995observation}, frequency standards~\cite{diddams2001optical, katori2003ultrastable}, quantum simulations~\cite{greiner2002quantum}, and cold collision studies~\cite{schmid2010dynamics}. These applications have made remarkable progress for alkali and alkaline earth atoms where atomic gases at ultralow temperatures can be readily prepared by laser cooling. This motivated researchers to cool atomic and molecular species with more complex internal structures, providing unique opportunities for investigating novel physical phenomena. 

Buffer gas cooling is a method to cool atomic or molecular species of interest to cryogenic temperatures via elastic collisions with cold, inert gas atoms, typically helium~\cite{hutzler2012buffer}. Since this cooling mechanism, unlike laser cooling, does not rely on the internal structure of the species to be cooled, buffer gas cooling is generically applicable to nearly any species that can be loaded into a buffer gas cell. Combined with beam production or trap loading, buffer gas cooling has expanded the scope of cold and ultracold physics. Notable examples include the search for physics beyond the Standard Model enhanced in sensitivity by a cryogenic molecular beam~\cite{acme2018improved} and the creation of a buffer-gas-cooled Bose-Einstein condensate~\cite{doret2009buffer}. Furthermore, buffer gas cooling serves as precooling for laser cooling and trapping of atomic or molecular species with internal complexity~\cite{shuman2010laser}. 

Despite the success of cold atomic physics, several atomic species of broad scientific interest have not yet been cooled to cryogenic temperatures. Among such atomic species, carbon is a particularly important element because of its essential role in natural sciences spanning biology, chemistry, and astronomy. As a method to obtain carbon atoms at cryogenic temperatures, laser cooling is still technically challenging because one-photon allowed electronic transitions of carbon lie in the vacuum ultraviolet (VUV) region. On the other hand, buffer gas cooling is a promising method to cool carbon atoms to the kelvin regime if technical difficulties in gas production and spectroscopic observation of ground-state carbon atoms can be overcome. 

Once cold samples of carbon are achieved, it will open up novel research opportunities. Carbon atoms in the kelvin regime could contribute to examining rich chemical reactions in the interstellar medium~\cite{henning1998carbon}. Utilizing two-photon allowed transitions, which are relatively weak but accessible with light sources in the deep ultraviolet (DUV) region, we can perform two-photon laser cooling to load the cold carbon atoms into a magnetic trap~\cite{sakamoto2022observation}. Trapped carbon atoms will not only be useful for precision spectroscopy but also provide a promising platform to pioneer cold collision studies or controlled chemistry of non-metallic atoms. 

In this Letter, we demonstrate buffer gas cooling of carbon atoms. The arrival time distribution of ablation-loaded carbon atoms at various buffer gas densities was obtained via in-cell, time-resolved spectroscopic measurements using two-photon excitation with broadband nanosecond pulses. We observed that the arrival time was delayed while maintaining the signal level as the buffer gas density increases. Comparisons with Monte Carlo trajectory simulations suggest that the local temperature of the carbon atoms we observed reach the temperature regime around 10~K within tens of microseconds after ablation at a buffer gas density of $4.7\times10^{15}~\pcc$. This confirms that buffer gas cooling combined with gas loading via ablation is a suitable method for producing slow carbon atoms, paving the way for interdisciplinary studies utilizing cold and ultracold atomic gases of carbon. 

\begin{figure}
\includegraphics[bb=0 0 243 232, scale=1, clip]{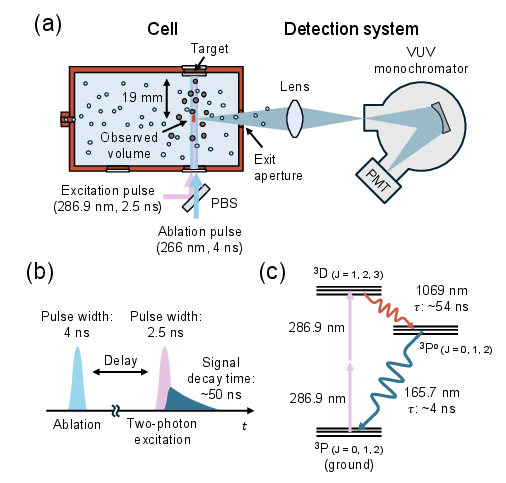}
\caption{(a) Schematic of the experimental setup (not to scale). Carbon atoms were produced by laser ablation of graphite. Broadband excitation pulses drove the $\ground$--$\excited$ two-photon transition of produced carbon atoms at 286.9~nm, followed by a two-step decay. The VUV photons associated with the decay were detected by the PMT through the exit aperture and a condenser lens. The observed volume was 19~mm away from the center of the target surface in the normal direction. (b) Timing chart for arrival time distribution measurements. (c) Relevant energy levels of carbon. The quantum number $J$ denotes the total angular momentum. The transition wavelength and lifetime are shown for the $\excited_{3}$--$\intermediate_{2}$ transition and the $\intermediate_{2}$--$\ground_{2}$ transition~\cite{NIST_ASD}.}
\label{fig1}
\end{figure}

A schematic of the experimental setup and relevant energy levels of carbon are given in Fig.~\ref{fig1}. The buffer gas cell was a copper box with an internal rectangular chamber of $40\times40~\mathrm{mm}^{2}$ cross-section and 70~mm length. The top of the cell was thermally anchored to the \ce{^{3}He} pot of a closed-cycle \ce{^{3}He} refrigerator. Here, we did not perform evaporative cooling of the liquid \ce{^{3}He}. The cell temperature was maintained at 3.2~K with no external heat load. The cell was equipped with UVFS windows on the front and back for optical access, a helium fill line on one side, and a circular exit aperture with a diameter of 5~mm on the other side for gas extraction. The cell was surrounded by two layers of aluminum radiation shields cooled by a two-stage pulse tube refrigerator. 

A helium buffer gas continuously flowed into the cell through a thin tube thermally anchored to the stages of the pulse tube refrigerator. We used a mass flow controller with a maximum mass flow of 10~SCCM to control the flow rate of the buffer gas. A good vacuum was maintained by a charcoal sorption pump bonded to the inner radiation shield and a turbomolecular pump attached to the bottom of the outer vacuum chamber. 

We loaded carbon atoms into the cell by laser ablation. The ablation target was a square plate of highly oriented pyrolytic graphite (HOPG). The target size was $10\times10~\mathrm{mm}^{2}$ cross-section and 1~mm thickness. Radiation at 266~nm from the fourth harmonic of a Q-switched Nd:YAG laser was irradiated onto the target, producing carbon atoms in the ground state~\cite{sakamoto2022observation}. The ablation laser had a repetition rate of 10~Hz, pulse width of 4~ns, and maximum pulse energy of 5.5~mJ. The spot size of the ablation laser beam on the target was 1.2~mm in $1/e^{2}$ diameter. As we have previously reported, setting a relatively low laser fluence so that the sample thermally sublimates is critical to produce ground-state carbon atoms in vacuum~\cite{sakamoto2022observation}. To alleviate the depletion of the number of produced atoms, we moved the ablation beam to a fresh spot on the sample at regular intervals. 

To detect ground-state carbon atoms, we excited them using linearly polarized, nanosecond pulses from a homebuilt tunable Ce:LiCAF laser, pumped by the fourth harmonic of a Q-switched Nd:YAG laser. The laser wavelength was tuned to the two-photon resonance of the $\ground$--$\excited$ transition at 286.9~nm. The excitation pulse had a duration of 2.5~ns, a repetition rate of 10~Hz, and a spectral width of 300~GHz. The broadband excitation pulse drove the $\ground$--$\excited$ two-photon transition of carbon atoms over the entire velocity distribution, followed by a two-step decay via the $\excited$--$\intermediate$ transition (lifetime: $\sim54~\mathrm{ns}$) and $\intermediate$--$\ground$ transition (lifetime: $\sim4~\mathrm{ns}$)~\cite{NIST_ASD}. The travel distance of the excited carbon atoms between two-photon excitation and photon emission was negligibly small. We introduced the excitation beam coaxially with the ablation beam using a polarizing beam splitter (PBS). Though two-photon excitation occurred over the entire beam of excitation pulses, the observed volume was limited to $200~\um\times200~\um\times2~\mathrm{mm}$, 19~mm away from the center of the target surface in the normal direction. Two-photon excitation-induced fluorescence at 165.7~nm was collected through the exit aperture of the cell in direction perpendicular to both the ablation and excitation beams. A 0.2-m monochromator equipped with a solar-blind photomultiplier tube (PMT) was used to detect the VUV photons. The fluorescence signal was recorded by a gated photon counter. We used a UVFS condenser lens with a solid angle of $2.5\times10^{-3}\times4\pi~\mathrm{sr}$ for detection. 

Since the fluorescence signal is proportional to the number of carbon atoms in the observed volume at the time of two-photon excitation, we can measure the arrival time distribution of carbon atoms by detecting the fluorescence as a function of the delay between the ablation pulse and excitation pulse. 
Without buffer gas, carbon atoms predominantly detected are those emitted normal to the target surface, and they are expected to travel ballistically. In the presence of buffer gas, on the other hand, we expect to observe carbon atoms that have reached the observed volume while being decelerated and changing the trajectories through collisions with the cold buffer gas atoms. Consequently, we expect the arrival time to shift towards the later side as the collision rate increases. 

\begin{figure}
\includegraphics[bb=0 0 243 170, scale=1, clip]{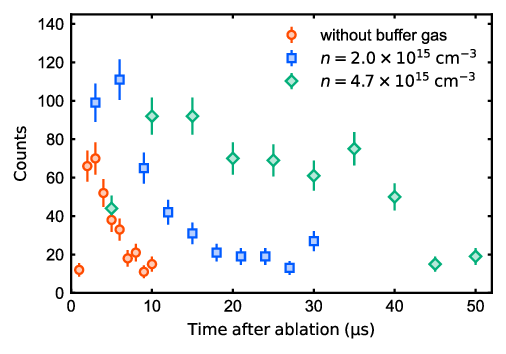}
\caption{Arrival time distributions of carbon atoms at various buffer gas densities. The cell temperature during the measurements was 6$\pm$1~K. At each data point, gated photon counting signals were integrated over 4800~shots. Error bars include statistical uncertainties from the photon shot noise.}
\label{fig2}
\end{figure}

Arrival time distributions of carbon atoms at various buffer gas densities are shown in Fig.~\ref{fig2}. The fluorescence was integrated over 4800~shots at each data point. The cell temperature during the measurements was 6$\pm$1~K. In the absence of buffer gas, the arrival time distribution peaks at around $3~\us$ with a full width at half maximum (FWHM) of $3~\us$, reflecting the initial velocity distribution of produced carbon atoms. Assuming a Maxwellian beam, the initial temperature is estimated to be 15000$\pm$2000~K, which is a few times higher than previously reported~\cite{sakamoto2022observation}. The discrepancy is due to the slightly higher ablation fluence that was set to maximize the signal level. In the presence of buffer gas, the distribution shows a clear dependence on the buffer gas density. For a buffer gas density of $2.0\times10^{15}~\pcc$, the peak was delayed to around $6~\us$ with a broader width of $9~\us$ (FWHM). For the maximum buffer gas density in this configuration of $4.7\times10^{15}~\pcc$, the peak was further delayed to around $15~\us$ with a width of $35~\us$ (FWHM). Here, we derived the buffer gas density from the flow rate, cell temperature, and exit aperture size~\cite{hutzler2012buffer}. 

Since carbon atoms colliding with more buffer gas atoms are expected to take longer flight time to reach the observed volume, these changes in the distributions qualitatively indicate the successful deceleration of carbon atoms by collisions with the buffer gas atoms. We note that the measurements with buffer gas are not intended to fit the waveforms with thermal distribution functions of certain temperatures, unlike the measurement without buffer gas. We will later discuss in detail the relationship between the measured arrival time distribution and the local temperature of the observed carbon atoms. 
Moreover, the signal level, proportional to the density of carbon atoms, was maintained as the buffer gas was introduced. Generally, an ambient gas modifies the nature of ablated particles, e.g., via cluster formation. Especially, in laser ablation of graphite, diverse clusters have been observed in the presence of buffer gas~\cite{straatsma2017production}, which potentially inhibits buffer gas cooling of carbon atoms. However, Fig.~\ref{fig2} shows that there was no signal reduction due to the presence of the buffer gas presumably because of the low carbon atom density that we discuss later. This confirms that buffer gas cooling combined with gas loading via laser ablation of graphite is a suitable method for producing slow-velocity carbon atoms. 

To evaluate the temperature of the buffer-gas-cooled species, a two-temperature model is typically used as follows~\cite{decarvalho1999buffer}:
\begin{align} 
T(N)=T_{\mathrm{b}} + (T'-T_{\mathrm{b}}) \times e^{(-\frac{N}{\kappa})},
\end{align}
where
\begin{align} 
\kappa \equiv \frac{(M+m)^{2}}{2Mm}.
\end{align}  
Here, $T'$ is the initial temperature of the species. $T(N)$ is the temperature of the species after $N$ collisions with buffer gas atoms. $T_{\mathrm{b}}$ is the buffer gas temperature. $M$ and $m$ are the masses of the species and the buffer gas atom, respectively, and $\kappa = 8/3$ in the present system. In addition, if the species is introduced via ablation, the buffer gas temperature temporarily increases and settles down to the cell temperature, typically on the millisecond timescale~\cite{PhysRevA.83.023418}. 

However, it is not appropriate to adopt the two-temperature model for describing the observed dynamics because the carbon atoms being decelerated are not in global thermal equilibrium, as described below. The momentum exchange rate between carbon atoms and helium atoms is relatively high compared with typical atomic or molecular species to be buffer-gas-cooled because the mass of a carbon atom is close to that of a helium atom. In addition, the density of produced carbon atoms is orders of magnitude lower than that of helium atoms. Consequently, the momentum decay rate of carbon atoms significantly exceeds the intraspecies momentum exchange rate, necessitating a description of the deceleration dynamics of carbon atoms through the tracking of collision events between carbon and helium atoms. 

We therefore performed Monte Carlo trajectory simulations and compared them with experimental results to evaluate the local temperature of the observed carbon atoms. We adopted the numerical model described in Ref.~\onlinecite{gantner2020buffer}. We traced the three-dimensional trajectory of a carbon atom on its random walk via elastic collisions with buffer gas atoms distributed uniformly in the cell. The ballistic flight distance of a carbon atom between collision events follows a probability distribution determined by the C--He collision cross-section and buffer gas density. The velocity vector of the carbon atom was modified by a collision with a buffer gas atom at a defined temperature. These processes were repeated until the carbon atom reached a defined boundary, giving the trajectory of the carbon atom. We terminated the trajectory calculation at that condition because carbon atoms colliding with the inner walls of the cell adhere there and do not contribute to the observation. 

\begin{figure*}
\includegraphics[bb=0 0 486 252, scale=1, clip]{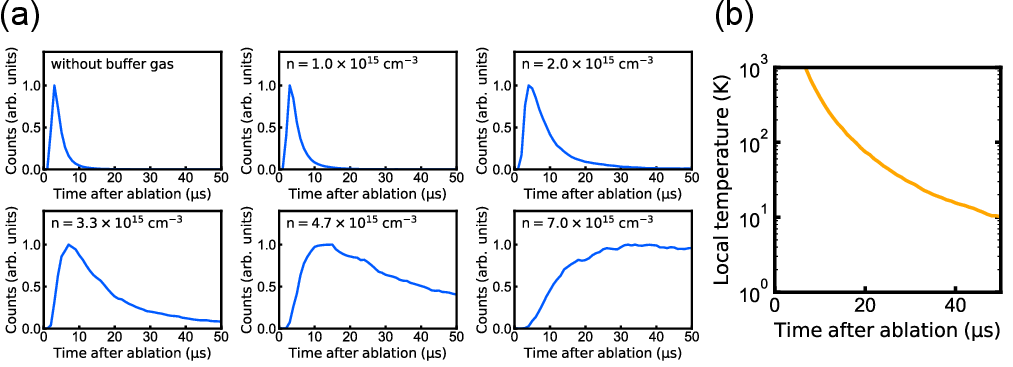}
\caption{(a) Simulated arrival time distributions 19~mm away from the ablated graphite surface at various buffer gas densities. Signal levels are normalized. The initial velocity distribution of carbon atoms is a Maxwellian beam at 15000~K for all the buffer gas densities. The buffer gas temperature is 6~K. We tune the C--He collision cross-section to $1.5\times10^{15}~\mathrm{cm}^{2}$ which best reproduces the experimental results. (b) Simulated temporal change in the local temperature of the observed carbon atoms at a buffer gas density of $4.7\times10^{15}~\pcc$.}
\label{fig3}
\end{figure*}

Fig.~\ref{fig3}(a) shows the calculated arrival time distribution at various buffer gas densities. We assumed that carbon atoms were emitted from the center of the target surface in the normal direction with initial velocities randomly picked up from a Maxwellian beam at 15000~K. We set the buffer gas temperature to 6~K. As discussed later in detail, considering the timescale of observation and the number of produced atoms, the heating of buffer gas atoms is negligible; thus, we fixed the buffer gas temperature. We tune the C--He collision cross-section to $1.5\times10^{15}~\mathrm{cm}^{2}$ which best reproduces the experimental results, while the hard-sphere model using the van der Waals radius gives a collision cross-section of $3.0\times10^{15}~\mathrm{cm}^{2}$. As pointed out in Ref.~\onlinecite{gantner2020buffer}, the collision cross-section given here cannot be directly compared to calculations based on more realistic models. 

The simulated distributions well capture the characteristics of the experimental results such as later peaks and longer durations as the buffer gas density increases. This confirms the validity of using the model to characterize the cooling process. To evaluate the achieved temperature of carbon atoms at the observed volume, we traced the velocity distribution of carbon atoms in the observed volume. Fig.~\ref{fig3}(b) shows the simulated temporal change in the local temperature of carbon atoms in the observed volume at a buffer gas density of $4.7\times10^{15}~\pcc$. The carbon atoms reach the temperature regime around 10~K within tens of microseconds after ablation, demonstrating that the carbon atoms we observed are actually buffer-gas cooled to cryogenic temperatures.

Generally, when the species of interest is introduced via ablation, the buffer gas temperature temporarily increases via collisions between produced hot particles and buffer gas atoms. However, the increase in the temperature of buffer gas atoms that decelerate carbon atoms is negligible, as described below. In the thermal ablation of graphite, hot particles including carbon atoms, molecules, and clusters are produced and heat the buffer gas. Among these produced particles, the carbon atom should be the fastest particle because it has the lightest mass. Therefore, it is reasonable to assume that carbon atoms are the first species to collide with cold buffer gas atoms and that the buffer gas atoms heated by other heavier species no longer contribute to the deceleration of the observed carbon atoms. In addition, the heating of the buffer gas by carbon atoms themselves is also negligibly small, as estimated below. The number of produced carbon atoms is estimated to be at most $1\times10^{9}$ considering the previously estimated value of $1.5\times10^{8}$~\cite{sakamoto2022observation} and the higher ablation laser fluence. For a buffer gas density of $4.7\times10^{15}~\pcc$, the volume where the carbon atoms are spread during the deceleration is estimated to be several~$\mathrm{cm}^{3}$, and the number of buffer gas atoms in the volume is more than 6~orders of magnitude larger than the number of carbon atoms. Under such a large difference in the number of atoms, the probability of heated buffer gas atoms re-colliding with carbon atoms is negligible, and when the heat is transferred to a sufficient fraction of the buffer gas, the increase in the buffer gas temperature is negligibly small. Based on the discussions above, the observed carbon atoms were cooled by a helium buffer gas at a fixed temperature of 6~K. 

The heating of the buffer gas is generally present on the microsecond timescale, as in Ref.~\onlinecite{PhysRevA.83.023418}, where a temporary increase in the buffer gas temperature by a few hundred kelvins is observed. The particularity of our system with respect to this general characteristics lies in the small number of carbon atoms. In Ref.~\onlinecite{PhysRevA.83.023418}, the number of produced YbF molecules is estimated to be $3\times10^{13}$, more than four orders of magnitude larger than the number of carbon atoms produced in this study. This presumably results from differences in the pulse energy of ablation pulses and the vapor pressure of the target. Moreover, since the YbF molecules are supposed to have completed thermalization with the buffer gas earlier than the time resolution in the study, other produced particles may have contributed to the heating of the buffer gas. 

In summary, we have demonstrated buffer gas cooling of carbon atoms. In-cell measurements of arrival time distributions of ablated carbon atoms at various buffer gas densities showed successful deceleration without any apparent reduction in atom number. Comparisons with Monte Carlo trajectory simulations suggest the observed carbon atoms locally reach the temperature regime around 10~K at a buffer gas density of $4.7\times10^{15}~\pcc$ within tens of microseconds after ablation. 

Buffer gas cooling of carbon atoms opens paths to new applications of cold atoms. Carbon atoms in the kelvin regime could provide a system for investigating rich chemical reactions involving neutral carbon in the laboratory, which is instrumental in understanding interstellar chemistry. Combined with laser cooling techniques, carbon atoms cooled to the microkelvin regime in the trap will develop cold collision studies and controlled chemistry of non-metallic atomic species, potentially impacting many fields ranging from astrophysics to biology. Furthermore, the precision of the unified atomic mass unit can be improved by precisely measuring the mass of a carbon atom using atom interferometry. 

Now that slow carbon atoms have been achieved, further characterization and cooling will be needed. To characterize the cooling process more accurately, time-resolved two-photon Doppler spectroscopy is useful. To simultaneously achieve the narrow linewidth and high two-photon excitation probability required for measurements, a single-mode pulsed laser in the DUV region with a microsecond-long duration is preferable~\cite{sakamoto2022observation}. Such a light source can also be utilized for two-photon Doppler-free frequency measurement of the laser cooling transition (i.e., the $\ground$--$\excited$ two-photon transition) and for two-photon laser cooling itself. A cryogenic beam of carbon atoms extracted from the cell, combined with subsequent two-photon laser cooling, allows us to load carbon atoms into a magnetic trap.

\begin{acknowledgments}
We would like to thank John M. Doyle for helpful discussions on the experimental setup for buffer gas cooling. This work was supported by Japan Science and Technology Agency Precursory Research for Embryonic Science and Technology (Grant No.~JPMJPR190B) and Ministry of Education, Culture, Sports, Science, and Technology Quantum Leap Flagship Program (Grant No.~JPMXS0118067246). T.~S. acknowledges support from the JSPS KAKENHI Grant No.~23KJ0592. 
\end{acknowledgments}

\bibliography{Reference}

\begin{thebibliography}{16}%
\makeatletter
\providecommand \@ifxundefined [1]{%
 \@ifx{#1\undefined}
}%
\providecommand \@ifnum [1]{%
 \ifnum #1\expandafter \@firstoftwo
 \else \expandafter \@secondoftwo
 \fi
}%
\providecommand \@ifx [1]{%
 \ifx #1\expandafter \@firstoftwo
 \else \expandafter \@secondoftwo
 \fi
}%
\providecommand \natexlab [1]{#1}%
\providecommand \enquote  [1]{``#1''}%
\providecommand \bibnamefont  [1]{#1}%
\providecommand \bibfnamefont [1]{#1}%
\providecommand \citenamefont [1]{#1}%
\providecommand \href@noop [0]{\@secondoftwo}%
\providecommand \href [0]{\begingroup \@sanitize@url \@href}%
\providecommand \@href[1]{\@@startlink{#1}\@@href}%
\providecommand \@@href[1]{\endgroup#1\@@endlink}%
\providecommand \@sanitize@url [0]{\catcode `\\12\catcode `\$12\catcode
  `\&12\catcode `\#12\catcode `\^12\catcode `\_12\catcode `\%12\relax}%
\providecommand \@@startlink[1]{}%
\providecommand \@@endlink[0]{}%
\providecommand \url  [0]{\begingroup\@sanitize@url \@url }%
\providecommand \@url [1]{\endgroup\@href {#1}{\urlprefix }}%
\providecommand \urlprefix  [0]{URL }%
\providecommand \Eprint [0]{\href }%
\providecommand \doibase [0]{https://doi.org/}%
\providecommand \selectlanguage [0]{\@gobble}%
\providecommand \bibinfo  [0]{\@secondoftwo}%
\providecommand \bibfield  [0]{\@secondoftwo}%
\providecommand \translation [1]{[#1]}%
\providecommand \BibitemOpen [0]{}%
\providecommand \bibitemStop [0]{}%
\providecommand \bibitemNoStop [0]{.\EOS\space}%
\providecommand \EOS [0]{\spacefactor3000\relax}%
\providecommand \BibitemShut  [1]{\csname bibitem#1\endcsname}%
\let\auto@bib@innerbib\@empty
\bibitem [{\citenamefont {Anderson}\ \emph {et~al.}(1995)\citenamefont
  {Anderson}, \citenamefont {Ensher}, \citenamefont {Matthews}, \citenamefont
  {Wieman},\ and\ \citenamefont {Cornell}}]{anderson1995observation}%
  \BibitemOpen
  \bibfield  {author} {\bibinfo {author} {\bibfnamefont {M.~H.}\ \bibnamefont
  {Anderson}}, \bibinfo {author} {\bibfnamefont {J.~R.}\ \bibnamefont
  {Ensher}}, \bibinfo {author} {\bibfnamefont {M.~R.}\ \bibnamefont
  {Matthews}}, \bibinfo {author} {\bibfnamefont {C.~E.}\ \bibnamefont
  {Wieman}},\ and\ \bibinfo {author} {\bibfnamefont {E.~A.}\ \bibnamefont
  {Cornell}},\ }\href@noop {} {\bibfield  {journal} {\bibinfo  {journal}
  {Science}\ }\textbf {\bibinfo {volume} {269}},\ \bibinfo {pages} {198}
  (\bibinfo {year} {1995})}\BibitemShut {NoStop}%
\bibitem [{\citenamefont {Diddams}\ \emph {et~al.}(2001)\citenamefont
  {Diddams}, \citenamefont {Udem}, \citenamefont {Bergquist}, \citenamefont
  {Curtis}, \citenamefont {Drullinger}, \citenamefont {Hollberg}, \citenamefont
  {Itano}, \citenamefont {Lee}, \citenamefont {Oates}, \citenamefont {Vogel}
  \emph {et~al.}}]{diddams2001optical}%
  \BibitemOpen
  \bibfield  {author} {\bibinfo {author} {\bibfnamefont {S.~A.}\ \bibnamefont
  {Diddams}}, \bibinfo {author} {\bibfnamefont {T.}~\bibnamefont {Udem}},
  \bibinfo {author} {\bibfnamefont {J.}~\bibnamefont {Bergquist}}, \bibinfo
  {author} {\bibfnamefont {E.}~\bibnamefont {Curtis}}, \bibinfo {author}
  {\bibfnamefont {R.}~\bibnamefont {Drullinger}}, \bibinfo {author}
  {\bibfnamefont {L.}~\bibnamefont {Hollberg}}, \bibinfo {author}
  {\bibfnamefont {W.~M.}\ \bibnamefont {Itano}}, \bibinfo {author}
  {\bibfnamefont {W.}~\bibnamefont {Lee}}, \bibinfo {author} {\bibfnamefont
  {C.}~\bibnamefont {Oates}}, \bibinfo {author} {\bibfnamefont
  {K.}~\bibnamefont {Vogel}}, \emph {et~al.},\ }\href@noop {} {\bibfield
  {journal} {\bibinfo  {journal} {Science}\ }\textbf {\bibinfo {volume}
  {293}},\ \bibinfo {pages} {825} (\bibinfo {year} {2001})}\BibitemShut
  {NoStop}%
\bibitem [{\citenamefont {Katori}\ \emph {et~al.}(2003)\citenamefont {Katori},
  \citenamefont {Takamoto}, \citenamefont {Pal'Chikov},\ and\ \citenamefont
  {Ovsiannikov}}]{katori2003ultrastable}%
  \BibitemOpen
  \bibfield  {author} {\bibinfo {author} {\bibfnamefont {H.}~\bibnamefont
  {Katori}}, \bibinfo {author} {\bibfnamefont {M.}~\bibnamefont {Takamoto}},
  \bibinfo {author} {\bibfnamefont {V.}~\bibnamefont {Pal'Chikov}},\ and\
  \bibinfo {author} {\bibfnamefont {V.}~\bibnamefont {Ovsiannikov}},\
  }\href@noop {} {\bibfield  {journal} {\bibinfo  {journal} {Phys. Rev. Lett.}\
  }\textbf {\bibinfo {volume} {91}},\ \bibinfo {pages} {173005} (\bibinfo
  {year} {2003})}\BibitemShut {NoStop}%
\bibitem [{\citenamefont {Greiner}\ \emph {et~al.}(2002)\citenamefont
  {Greiner}, \citenamefont {Mandel}, \citenamefont {Esslinger}, \citenamefont
  {H{\"a}nsch},\ and\ \citenamefont {Bloch}}]{greiner2002quantum}%
  \BibitemOpen
  \bibfield  {author} {\bibinfo {author} {\bibfnamefont {M.}~\bibnamefont
  {Greiner}}, \bibinfo {author} {\bibfnamefont {O.}~\bibnamefont {Mandel}},
  \bibinfo {author} {\bibfnamefont {T.}~\bibnamefont {Esslinger}}, \bibinfo
  {author} {\bibfnamefont {T.~W.}\ \bibnamefont {H{\"a}nsch}},\ and\ \bibinfo
  {author} {\bibfnamefont {I.}~\bibnamefont {Bloch}},\ }\href@noop {}
  {\bibfield  {journal} {\bibinfo  {journal} {Nature}\ }\textbf {\bibinfo
  {volume} {415}},\ \bibinfo {pages} {39} (\bibinfo {year} {2002})}\BibitemShut
  {NoStop}%
\bibitem [{\citenamefont {Schmid}\ \emph {et~al.}(2010)\citenamefont {Schmid},
  \citenamefont {H{\"a}rter},\ and\ \citenamefont
  {Denschlag}}]{schmid2010dynamics}%
  \BibitemOpen
  \bibfield  {author} {\bibinfo {author} {\bibfnamefont {S.}~\bibnamefont
  {Schmid}}, \bibinfo {author} {\bibfnamefont {A.}~\bibnamefont {H{\"a}rter}},\
  and\ \bibinfo {author} {\bibfnamefont {J.~H.}\ \bibnamefont {Denschlag}},\
  }\href@noop {} {\bibfield  {journal} {\bibinfo  {journal} {Phys. Rev. Lett.}\
  }\textbf {\bibinfo {volume} {105}},\ \bibinfo {pages} {133202} (\bibinfo
  {year} {2010})}\BibitemShut {NoStop}%
\bibitem [{\citenamefont {Hutzler}\ \emph {et~al.}(2012)\citenamefont
  {Hutzler}, \citenamefont {Lu},\ and\ \citenamefont
  {Doyle}}]{hutzler2012buffer}%
  \BibitemOpen
  \bibfield  {author} {\bibinfo {author} {\bibfnamefont {N.~R.}\ \bibnamefont
  {Hutzler}}, \bibinfo {author} {\bibfnamefont {H.-I.}\ \bibnamefont {Lu}},\
  and\ \bibinfo {author} {\bibfnamefont {J.~M.}\ \bibnamefont {Doyle}},\
  }\href@noop {} {\bibfield  {journal} {\bibinfo  {journal} {Chem. Rev.}\
  }\textbf {\bibinfo {volume} {112}},\ \bibinfo {pages} {4803} (\bibinfo {year}
  {2012})}\BibitemShut {NoStop}%
\bibitem [{\citenamefont {{ACME Collaboration}}(2018)}]{acme2018improved}%
  \BibitemOpen
  \bibfield  {author} {\bibinfo {author} {\bibnamefont {{ACME
  Collaboration}}},\ }\href@noop {} {\bibfield  {journal} {\bibinfo  {journal}
  {Nature}\ }\textbf {\bibinfo {volume} {562}},\ \bibinfo {pages} {355}
  (\bibinfo {year} {2018})}\BibitemShut {NoStop}%
\bibitem [{\citenamefont {Doret}\ \emph {et~al.}(2009)\citenamefont {Doret},
  \citenamefont {Connolly}, \citenamefont {Ketterle},\ and\ \citenamefont
  {Doyle}}]{doret2009buffer}%
  \BibitemOpen
  \bibfield  {author} {\bibinfo {author} {\bibfnamefont {S.~C.}\ \bibnamefont
  {Doret}}, \bibinfo {author} {\bibfnamefont {C.~B.}\ \bibnamefont {Connolly}},
  \bibinfo {author} {\bibfnamefont {W.}~\bibnamefont {Ketterle}},\ and\
  \bibinfo {author} {\bibfnamefont {J.~M.}\ \bibnamefont {Doyle}},\ }\href@noop
  {} {\bibfield  {journal} {\bibinfo  {journal} {Phys. Rev. Lett.}\ }\textbf
  {\bibinfo {volume} {103}},\ \bibinfo {pages} {103005} (\bibinfo {year}
  {2009})}\BibitemShut {NoStop}%
\bibitem [{\citenamefont {Shuman}\ \emph {et~al.}(2010)\citenamefont {Shuman},
  \citenamefont {Barry},\ and\ \citenamefont {DeMille}}]{shuman2010laser}%
  \BibitemOpen
  \bibfield  {author} {\bibinfo {author} {\bibfnamefont {E.~S.}\ \bibnamefont
  {Shuman}}, \bibinfo {author} {\bibfnamefont {J.~F.}\ \bibnamefont {Barry}},\
  and\ \bibinfo {author} {\bibfnamefont {D.}~\bibnamefont {DeMille}},\
  }\href@noop {} {\bibfield  {journal} {\bibinfo  {journal} {Nature}\ }\textbf
  {\bibinfo {volume} {467}},\ \bibinfo {pages} {820} (\bibinfo {year}
  {2010})}\BibitemShut {NoStop}%
\bibitem [{\citenamefont {Henning}\ and\ \citenamefont
  {Salama}(1998)}]{henning1998carbon}%
  \BibitemOpen
  \bibfield  {author} {\bibinfo {author} {\bibfnamefont {T.}~\bibnamefont
  {Henning}}\ and\ \bibinfo {author} {\bibfnamefont {F.}~\bibnamefont
  {Salama}},\ }\href@noop {} {\bibfield  {journal} {\bibinfo  {journal}
  {Science}\ }\textbf {\bibinfo {volume} {282}},\ \bibinfo {pages} {2204}
  (\bibinfo {year} {1998})}\BibitemShut {NoStop}%
\bibitem [{\citenamefont {Sakamoto}\ and\ \citenamefont
  {Yoshioka}(2022)}]{sakamoto2022observation}%
  \BibitemOpen
  \bibfield  {author} {\bibinfo {author} {\bibfnamefont {T.}~\bibnamefont
  {Sakamoto}}\ and\ \bibinfo {author} {\bibfnamefont {K.}~\bibnamefont
  {Yoshioka}},\ }\href {https://doi.org/10.1103/PhysRevA.106.052808} {\bibfield
   {journal} {\bibinfo  {journal} {Phys. Rev. A}\ }\textbf {\bibinfo {volume}
  {106}},\ \bibinfo {pages} {052808} (\bibinfo {year} {2022})}\BibitemShut
  {NoStop}%
\bibitem [{\citenamefont {Kramida}\ \emph {et~al.}(2023)\citenamefont
  {Kramida}, \citenamefont {{Yu.~Ralchenko}}, \citenamefont {Reader},\ and\
  \citenamefont {{NIST ASD Team}}}]{NIST_ASD}%
  \BibitemOpen
  \bibfield  {author} {\bibinfo {author} {\bibfnamefont {A.}~\bibnamefont
  {Kramida}}, \bibinfo {author} {\bibnamefont {{Yu.~Ralchenko}}}, \bibinfo
  {author} {\bibfnamefont {J.}~\bibnamefont {Reader}},\ and\ \bibinfo {author}
  {\bibnamefont {{NIST ASD Team}}},\ }\href@noop {} {}\bibinfo {howpublished}
  {{NIST Atomic Spectra Database (ver. 5.11), [Online]. Available:
  {\tt{https://physics.nist.gov/asd}} [2024, May 31]. National Institute of
  Standards and Technology, Gaithersburg, MD.}} (\bibinfo {year}
  {2023})\BibitemShut {NoStop}%
\bibitem [{\citenamefont {Straatsma}\ \emph {et~al.}(2017)\citenamefont
  {Straatsma}, \citenamefont {Fabrikant}, \citenamefont {Douberly},\ and\
  \citenamefont {Lewandowski}}]{straatsma2017production}%
  \BibitemOpen
  \bibfield  {author} {\bibinfo {author} {\bibfnamefont {C.}~\bibnamefont
  {Straatsma}}, \bibinfo {author} {\bibfnamefont {M.}~\bibnamefont
  {Fabrikant}}, \bibinfo {author} {\bibfnamefont {G.}~\bibnamefont
  {Douberly}},\ and\ \bibinfo {author} {\bibfnamefont {H.}~\bibnamefont
  {Lewandowski}},\ }\href@noop {} {\bibfield  {journal} {\bibinfo  {journal}
  {J. Chem. Phys.}\ }\textbf {\bibinfo {volume} {147}} (\bibinfo {year}
  {2017})}\BibitemShut {NoStop}%
\bibitem [{\citenamefont {deCarvalho}\ \emph {et~al.}(1999)\citenamefont
  {deCarvalho}, \citenamefont {Doyle}, \citenamefont {Friedrich}, \citenamefont
  {Guillet}, \citenamefont {Kim}, \citenamefont {Patterson},\ and\
  \citenamefont {Weinstein}}]{decarvalho1999buffer}%
  \BibitemOpen
  \bibfield  {author} {\bibinfo {author} {\bibfnamefont {R.}~\bibnamefont
  {deCarvalho}}, \bibinfo {author} {\bibfnamefont {J.~M.}\ \bibnamefont
  {Doyle}}, \bibinfo {author} {\bibfnamefont {B.}~\bibnamefont {Friedrich}},
  \bibinfo {author} {\bibfnamefont {T.}~\bibnamefont {Guillet}}, \bibinfo
  {author} {\bibfnamefont {J.}~\bibnamefont {Kim}}, \bibinfo {author}
  {\bibfnamefont {D.}~\bibnamefont {Patterson}},\ and\ \bibinfo {author}
  {\bibfnamefont {J.~D.}\ \bibnamefont {Weinstein}},\ }\href@noop {} {\bibfield
   {journal} {\bibinfo  {journal} {Eur. Phys. J. D}\ }\textbf {\bibinfo
  {volume} {7}},\ \bibinfo {pages} {289} (\bibinfo {year} {1999})}\BibitemShut
  {NoStop}%
\bibitem [{\citenamefont {Skoff}\ \emph {et~al.}(2011)\citenamefont {Skoff},
  \citenamefont {Hendricks}, \citenamefont {Sinclair}, \citenamefont {Hudson},
  \citenamefont {Segal}, \citenamefont {Sauer}, \citenamefont {Hinds},\ and\
  \citenamefont {Tarbutt}}]{PhysRevA.83.023418}%
  \BibitemOpen
  \bibfield  {author} {\bibinfo {author} {\bibfnamefont {S.~M.}\ \bibnamefont
  {Skoff}}, \bibinfo {author} {\bibfnamefont {R.~J.}\ \bibnamefont
  {Hendricks}}, \bibinfo {author} {\bibfnamefont {C.~D.~J.}\ \bibnamefont
  {Sinclair}}, \bibinfo {author} {\bibfnamefont {J.~J.}\ \bibnamefont
  {Hudson}}, \bibinfo {author} {\bibfnamefont {D.~M.}\ \bibnamefont {Segal}},
  \bibinfo {author} {\bibfnamefont {B.~E.}\ \bibnamefont {Sauer}}, \bibinfo
  {author} {\bibfnamefont {E.~A.}\ \bibnamefont {Hinds}},\ and\ \bibinfo
  {author} {\bibfnamefont {M.~R.}\ \bibnamefont {Tarbutt}},\ }\href
  {https://doi.org/10.1103/PhysRevA.83.023418} {\bibfield  {journal} {\bibinfo
  {journal} {Phys. Rev. A}\ }\textbf {\bibinfo {volume} {83}},\ \bibinfo
  {pages} {023418} (\bibinfo {year} {2011})}\BibitemShut {NoStop}%
\bibitem [{\citenamefont {Gantner}\ \emph {et~al.}(2020)\citenamefont
  {Gantner}, \citenamefont {Koller}, \citenamefont {Wu}, \citenamefont
  {Rempe},\ and\ \citenamefont {Zeppenfeld}}]{gantner2020buffer}%
  \BibitemOpen
  \bibfield  {author} {\bibinfo {author} {\bibfnamefont {T.}~\bibnamefont
  {Gantner}}, \bibinfo {author} {\bibfnamefont {M.}~\bibnamefont {Koller}},
  \bibinfo {author} {\bibfnamefont {X.}~\bibnamefont {Wu}}, \bibinfo {author}
  {\bibfnamefont {G.}~\bibnamefont {Rempe}},\ and\ \bibinfo {author}
  {\bibfnamefont {M.}~\bibnamefont {Zeppenfeld}},\ }\href@noop {} {\bibfield
  {journal} {\bibinfo  {journal} {J. Phys. B}\ }\textbf {\bibinfo {volume}
  {53}},\ \bibinfo {pages} {145302} (\bibinfo {year} {2020})}\BibitemShut
  {NoStop}%
\end{thebibliography}%
\bibliographystyle{apsrev4-2}

\end{document}